\documentstyle[twoside,amssymb,12pt]{article}

\newcommand{\nc}{\newcommand}
\nc{\la}{\lambda} \nc{\alf}{\alpha}
\nc{\tht}{\theta}  \nc{\be}{\beta}  \nc{\eps}{\epsilon} 
\nc{\ga}{\gamma}  \nc{\De}{\Delta}  \nc{\G}{\Gamma}  \nc{\vphi}{\varphi}
\nc{\de}{\delta} \nc{\si}{\sigma}  \nc{\ka}{\kappa}   \nc{\Si}{\Sigma} 
\nc{\om}{\omega}  \nc{\qq}{\quad\quad}                \nc{\Om}{\Omega}
\nc{\nf}{\infty}   \nc{\dl}{\mathop{\smash{\cal L}}} 
\nc{\ra}{\rightarrow}    \nc{\ol}{\overline}        \nc{\und}{\underline} 
\nc{\beq}{\begin{equation}}  \nc{\eeq}{\end{equation}}  \nc{\pt}{\partial}  
   \nc{\dst}{\displaystyle}  
\nc{\nnb}{\nonumber}    \nc{\bs}{\backslash}        \nc{\mb}{\mathbb}   
\nc{\sn}{{\rm sn}\,} \nc{\cn}{{\rm cn}\,}     \nc{\dn}{{\rm dn}\,}
\nc{\ti}{\tilde}   \nc{\wti}{\widetilde}   \nc{\h}{\hat}  \nc{\wh}{\widehat}

\newcounter{muni}
\newenvironment{remunerate}{\begin{list}{{\rm \arabic{muni}.}}
{\usecounter{muni}
\setlength{\leftmargin}{0pt}\setlength{\itemindent}{38pt}}}{\end{list}}

\nc{\brm}{\begin{remunerate}}   \nc{\erm}{\end{remunerate}}

\nc{\stg}{\mathop{\smash{*}}}
\nc{\st}{\mathop{\smash{\delta}}}
\nc{\barr}{\begin{array}}   \nc{\earr}{\end{array}}   \nc{\dg}{\dagger}
\nc{\mtvb}{\mathversion{bold}}   \nc{\mtvn}{\mathversion{normal}}

\title{\bf One-loop renormalizability of all\\ 2d dimensional Poisson-Lie
$\si$-models} \author{Ctirad KLIM\v CIK
\thanks{\noindent Institut de Math\'ematiques de Luminy, CNRS UPR 9016, 
Case 930, 163 Avenue de Luminy, 13288 Marseille Cedex 9, France}  
\and Galliano VALENT
\thanks{\noindent Centre de Physique Th\'eorique, CNRS Luminy, Case 907, 
13288 Marseille Cedex 9, France.}
\thanks{\noindent Laboratoire de Physique Th\'eorique et des Hautes Energies, 
CNRS UMR 7859,~Universit\'e Paris 7,
2 Place Jussieu, 75251 Paris Cedex 05, France.}}

\date{April 8, 2003}
\begin{document}
\maketitle

\vspace{-8cm}
\begin{flushright}
CPT 03-4530 $\hspace{3mm}$ \\
IML 03-03 $\hspace{3mm}$ \\
LPTHE 03-11 $\hspace{3mm}$ \\
hep-th/0304053 $\hspace{3mm}$  \\
April 2003 $\hspace{3mm}$ 
\end{flushright}
%\vskip 1.0truecm

\vspace{6cm}

\begin{abstract}
We perform a systematic study of the one-loop renormalizability of all 
Poisson-Lie T-dualizable $\si$-models with two-dimensional targets. We 
show that whatever Drinfeld double and whatever matrix of coupling 
constants we consider the corresponding $\si$-model is always one-loop 
renormalizable in the strict field theoretical sense. Moreover, in all 
cases, the RG flow in the space of the coupling 
constants is compatible with the Poisson-Lie T-duality.
\end{abstract}  

\newpage
\section{Introduction}
Poisson-Lie T-duality \cite{ks1,K} is the generalization of the 
world-sheet Abelian T-duality in string theory \cite{ky,ss} and of 
the traditional non-Abelian T-duality \cite{ft,fj,oq}. It works also 
in situations when a T-dualizable $\si$-model does not possess a 
(non)Abelian isometry but only a weaker property called 
Poisson-Lie symmetry.
                                                                     
The Poisson-Lie T-dualizable $\si$-models (or, in what follows, the PLT 
$\si$-models) are specified by the choice of a Drinfeld double and by the 
$(n\times n)$-matrix of coupling constants where $n$ is the half of the 
dimension of the double. From the point of view of classical field theory, 
the $\si$-model and its dual are related by a canonical transformation 
\cite{ks2,S1,A}; they are therefore dynamically equivalent systems.

The quantum status of the Poisson-Lie T-duality remains a challenging 
problem. We stress at this point that the word ``quantum" does not 
necessarily suppose that a conformal symmetry is to be required. We 
shall simply study the duality from the point of two-dimensional 
field-theory. Whereas for the semi-abelian case the one-loop quantum 
equivalence of T-dual models has been established for a wide class of 
models \cite{cv}, the same problem for the non-abelian case is far 
more difficult. The first steps in this direction were undertaken 
in \cite{S2,bf}, where the respective authors established the one-loop 
renormalizability of certain Poisson-Lie T-dualizable $\si$-models 
and the compatibility of the RG flow with the Poisson-Lie T-duality. 
They did it for a few low-dimensional Poisson-Lie targets and 
particular choices of the coupling constants.

In this paper, we would like to perform a more systematic study of 
this issue. However, we miss a classification of all PLT $\si$-models 
since it would require a preliminary classification of all the 
Drinfeld doubles (indeed, the latter project 
seems as hopeless as the classification of all Lie algebras). On the 
other hand, low dimensional doubles WERE classified in \cite{hs1,jra,hs2}. 
Therefore we consider, e. g., {\em all} existing two-dimensional 
Poisson-Lie T-dualizable targets. We do it and show that they are 
one-loop renormalizable and that the RG flow in the space of 
coupling constants is always compatible with the Poisson-Lie T-duality.

\section{Generic 2 dimensional Poisson-Lie models}
\subsection{Four dimensional Drinfeld doubles}
A Drinfeld double $\,{\cal D}\,$ is a Lie algebra  
with generators $\,T_i,\ i=1,\ldots n$ and  
$\,\wti{T}^i,\ i=1,\ldots n,$ equipped with a symmetric and ad-invariant 
non-degenerate bilinear form $\langle\cdot,\cdot\rangle$ such that
\beq\label{dr1}
\langle T_i,T_j\rangle=0,\qq\langle\wti{T}^i,\wti{T}^j\rangle=0,\qq 
\langle T_i,\wti{T}^j\rangle=\langle\wti{T}^j,T_i\rangle=\de_i^j.\eeq 
It is moreover required that the linear subspaces 
${\cal G}\equiv Span(T_i)$ and $\ti{\cal G}\equiv Span(\ti{T}^i)$ are 
respectively the subalgebras of ${\cal D}.$

As shown by \cite{hs1} all the four-dimensional non-isomorphic 
Drinfeld doubles, denoted as $\,{\cal D}(\rho,\nu),$ can be written  
\beq\label{dr2}\left\{\barr{lll}
[T_1,T_2]=\rho\,T_2\qq & \qq [T_1,\ti{T}^1]=0 & \qq [T_1,\ti{T}^2]=
-\rho\,\ti{T}^2 \\[4mm]
[\ti{T}^1,\ti{T}^2]=\nu\,\ti{T}^2\qq & \qq [T_2,\ti{T}^1]=\nu\,T_2 & 
\qq [T_2,\ti{T}^2]=\rho\,\ti{T}^1-\nu\,T_1\earr\right.\eeq
and are of {\em three} non-isomorphic types:
\brm
\item The fully abelian double $\,{\cal D}(0,0).$
\item The semi-abelian double $\,{\cal D}(1,0).$
\item The non-abelian double $\,{\cal D}(1,\nu)\,$ with $\,\nu\neq 0.$\erm
Let us construct the Poisson-Lie $\si$-models out of these doubles.

\subsection{The Poisson-Lie models and their T-duals}
A general Poisson-Lie T-dualizable $\si$-model has for action \cite{ks1,ks2}
\beq\label{pl1}
\int\,(R(g)+\Pi)^{-1}_{ij} \,(\pt_+g\,g^{-1})^i\,(\pt_-g\,g^{-1})^j,\eeq
where
\beq  \pt_{\pm}=\pt_{\tau}\pm\pt_{\si}\eeq
and $\tau$ and $\si$ are the ``time " and ``space" coordinates on the 
world-sheet. The model (\ref{pl1}) lives on a group manifold $G$ 
(corresponding to the Lie algebra ${\cal G}$) and it is parametrized 
by a set of coupling constants assembled into the matrix $R.$

In order to determine the $g$-dependent matrix $\,\Pi(g)\,$ one 
first defines a triplet of matrices as follows
\beq\label{pl11}
{\rm Ad}\,(g^{-1})T_i\equiv g^{-1}T_ig=a(g)_i^{~l}\,T_l,\qq
{\rm Ad}\,(g^{-1})\wti{T}^i=b(g)^{il}\,T_l+d(g)^i_{~l}\,\wti{T}^l.\eeq
Then
\beq\Pi(g)=b(g)\,a^{-1}(g).\eeq
Note that the matrices $a,b,d$ are defined by the adjoint action of  
$g\in G$ on the Lie algebra ${\cal D}.$

In our two-dimensional case, we may realize the group $G$ in the 
matrix way as follows
\beq
g=\left(\barr{cc} e^{\rho\chi} & \rho\tht\\[4mm] 0 & 1\earr\right)\eeq
and the generators $T_i$, $\wti{T}^i$ are given by
\beq
T_1=\left(\barr{cc}  \rho & 0\\[4mm] 0 & 0\earr\right),\quad
T_2=\left(\barr{cc}  0 & \rho \\[4mm] 0 & 0\earr\right),\quad
 \ti T^1=\left(\barr{cc}  0 & 0\\[4mm] 0 & \nu\earr\right),\quad
\ti T^2=\left(\barr{cc}  0 & 0 \\[4mm] -\nu & 0\earr\right).
\eeq
Thus we have for the matrices
\beq\label{pll11}
a(g)=\left(\barr{cc} 1 & \rho\tht e^{-\rho\chi}\\[4mm]
 0 & e^{-\rho\chi}\earr\right),\qq\quad 
b(g)=\left(\barr{cc} 0 & -\nu\tht e^{-\rho\chi}\\[4mm] 
\nu\tht & \rho\nu\tht^2 e^{-\rho\chi}\earr\right),\eeq
which lead to
\beq\label{pl12}
\Pi(g)=b(g)\,a^{-1}(g)\quad\Rightarrow\quad
\Pi=\left(\barr{cc} 0 & -\nu\tht \\[4mm] 
\nu\tht & 0\earr\right).\eeq
The matrix $\,R\,$ is
\beq\label{pl2}
R=\left(\barr{cc} x & y \\[4mm] z & w\earr\right),
\qq\qq \de\equiv\det R=xw-yz\neq 0,\eeq
and the right-invariant vielbein
\beq\label{pl3}
dg\,g^{-1}=d\chi\,T_1+(d\tht-\rho\tht d\chi)\,T_2.\eeq

Then we obtain from the action (\ref{pl1}) the 
resulting target space metric \cite{K}, \cite{hs1}
\beq\label{pl5}
G=\frac b{\De}\,d\chi^2+\frac x{\De}\,d\tht^2
-2\frac a{\De}\,d\chi\,d\tht,
\qq\De=(\nu\tht)^2-(y-z)(\nu\tht)+\de,\eeq
with the definitions
\beq\label{pl6}
a=x(\rho\tht)+\frac{y+z}{2},\qq b=x(\rho\tht)^2+(y+z)(\rho\tht)+w,\eeq
and the torsion potential
\beq\label{pl7}
\frac{\rho\tht-(y-z)/2}{\De}\ d\chi\wedge d\tht.\eeq
Since the target space is two-dimensional, the torsion 3-form vanishes. 

The T-dualized model lives on the dual group target $\wti{G}$ and 
its action reads \cite{ks1,ks2}
\beq\label{dual}
 \int {(\wti R+\wti\Pi(\ti g))^{-1}}^{ij}
\,(\pt_+\ti g\,\ti g^{-1})_i\,(\pt_-\ti g\,\ti g^{-1})_j,
\eeq
where $\wti R=R^{-1}$ and the $\ti g$ dependent matrix is given by
\beq\wti{\Pi}(\ti{g})=\ti{b}(\ti{g})\ti{a}^{-1}(\ti{g}),\eeq
where the dual matrices $\ti a,\ti b,\ti d$ are defined similarly by
\beq 
{\rm Ad}_{\ti g^{-1}}\ti T^i= \ti a(\ti g)^i_{~l}\ti T^l,\qq
{\rm Ad}_{\ti g^{-1}} T_i=\ti b(\ti g)_{il}\ti T^l+\ti d(\ti g)_i^{~l}T_l.
\eeq 
The elements $\ti g$ of the dual group $\wti G$ are of the form
\beq \ti g=\left(\barr{cc} 0&0\\[4mm] -\nu\tht & e^{ \nu\chi}\earr\right)\eeq
and one can check that the matrices$\,\ti{a}(\ti{g}),\,\ti{b}(\ti{g})\,$ are 
obtained from $\,a(g),\,b(g)\,$ by the interchange of the parameters 
$\,\rho\,$ and $\,\nu.$ This leads to
\beq\label{dual2}
\wti \Pi(\ti g)=\left(\barr{cc} 0 & -\rho\tht\\[4mm]\rho\tht & 0\earr\right).
\eeq
 
Finally, the vielbein is now
\beq\label{dual4}
d\ti{g}\,\ti{g}^{-1}=d\chi\,\ti{T}_1+(d\tht-\nu\tht\,d\chi)\,\ti{T}_2.
\eeq
Inserting all these quantities in the T-dualized Poisson-Lie model 
(\ref{dual})gives the same target space metric  (\ref{pl5}) in 
which $\,R\,$ is transformedinto $\,\wti{R}\,$ and $\,\rho\,$ and 
$\,\nu\,$ get exchanged.

From the previous considerations one can check that the abelian model and its 
T-dual partner are both flat, so we will not consider this trivial possibility
and we will take $\rho=1$ in what follows.

\subsection{Isometry and form invariance of the metric}
These metrics have the obvious Killing $\wti{K}=\pt_{\chi},$ with 
dual 1-form 
\beq\label{fi3}
K=\frac{a^2+\ga^2}{x\De}\,d\chi-\frac a{\De}\,d\tht.\eeq

Besides this isometry, the left group action induces a two 
parameters group of transformation $\,G_L\,$ of the coordinates 
and of the parameters 
\beq\label{fi1}\left\{\barr{l}
\wh{\tht}=\si\tht-\tau,\qq\wh{\chi}=\chi+\ln\si,\qq\wh{x}=x,\\[4mm]
\wh{y}=\si y+\tau(x-1),\qq \wh{z}=\si z+\tau(x+1),\\[4mm]
\wh{w}=\si^2 w+\si\tau(y+z)+\tau^2 x,
\earr\right.\qq \si>0,\ \ \tau\in{\mb R}.\eeq
Using the relations
\beq\label{fi2}
\wh{a}=\si a,\qq\qq \wh{b}=\si^2 b, \qq\qq\wh{\De}=\si^2\De,\eeq
it is easy to check that $\,G_L\,$ leaves the metric invariant in the 
sense that
\[G(\chi,\tht,R)=G(\wh{\chi},\wh{\tht},\wh{R}).\]
The existence of this group implies that we can get rid 
of two parameters in the matrix $\,R:$ we can always take $\,z=y\,$ and
$\,w=x.$  Indeed, let us consider a set of the parameters $\,x,\,y,\,z,\,w\,$
such that  neither $\,z=y\,$ nor $w=x.$ Using the transformations laws given
above, if we  take for parameters
\[\tau=\si\,\frac{(y-z)}{2},\qq\qq
\si=x\left(\frac{t^2}{4}+\ga^2\right)^{-1/2},\qq t=(x+1)y+(x-1)z,\]
we can ensure simultaneously the relations $\,\wh{z}=\wh{y}\,$ and
$\wh{w}=\wh{x}.$  Notice that we have anticipated the positivity restriction
on $x$   which comes out from the riemannian character of the metric (see
(\ref{riem})).  In what follows we will therefore take
\beq\label{R1}
R=\left(\barr{cc} x & y \\[4mm] y & x\earr\right),\qq x>0,\quad 
\ga^2\equiv x^2-y^2>0,\quad y\in{\mb R}.\eeq
                                                                               
For the semi-abelian model, corresponding to $\,{\cal D}(1,0),$ the 
lagrangian is
\[B_{ij}\,(\pt_+g\,g^{-1})^i\,(\pt_-g\,g^{-1})^j,\qq B=R^{-1},\]
where $R$ is given in (\ref{R1}). This metric is flat, so there is no
renormalization of the parameters. Its T-dual model is not flat, but 
as shown in \cite{cv}, it is flat up to some diffeomorphism, and the 
one-loop equivalence between the T-dual pair of models is preserved. 

So the remaining open problem is the non-abelian
model: since $\,\nu\neq 0,$ appropriate scalings of $\,R\,$ and of the 
coordinates allow to take $\,\nu=1.$ As observed in \cite{hs1} the full 
family of  Drinfeld doubles $\,{\cal D}(1,\nu)\,$ gives rise to a single
$\,\si$-model  with  $\,\nu=1,$ the $\,GL(2,{\mb R})\,$ model of \cite{K}.
 
The actual form of the metric is now
\beq\label{nmet1}
G=\frac b{\De}\,d\chi^2+\frac x{\De}\,d\tht^2
-2\frac a{\De}\,d\chi\,d\tht,\eeq
with the new functions
\beq\label{nmet2}
a=x\tht+y,\qq b=x\tht^2+2y\tht+x=\frac{a^2+\ga^2}{x},
\qq \De=\tht^2+\ga^2,\quad \ga^2=x^2-y^2.\eeq
Its T-dual partner is obtained by the change of parameters
\beq\label{nmet3}
x\ \ \to\ \ \wti{x}=\frac x{x^2-y^2},\qq\quad 
y\ \ \to\ \ \wti{y}=-\frac y{x^2-y^2}.\eeq

\subsection{Geometric aspects}
Let us check the Riemannian character of the metric (\ref{nmet1}). This 
is best done through a calculation of the vielbein 
\beq\label{pl8}
G=(e_1)^2+(e_2)^2,\qq e_1=\frac{\ga}{\sqrt{x\De}}\,d\chi,\qq 
e_2=\frac{x\,d\tht-a\,d\chi}{\sqrt{x\De}},\eeq
The metric will be riemannian iff
\beq\label{riem}\ga^2>0,\qq\qq x>0.\eeq
It follows that $\,\De\,$ never vanishes. In the sequel we will 
assume that these conditions hold.

Another interesting point is to find the coordinates which do exhibit 
the conformally flat character of this metric. Taking for 
these coordinates 
\beq\label{ff2}
U=\chi-\frac 12\ln(a^2(\tht)+\ga^2),\qq\quad
V=\arctan\left(\frac{a(\tht)}{\ga}\right),\qq a(\tht)=x\tht+y,\eeq
the metric $\,G\,$ becomes
\beq\label{ff1}
G=\frac{x\ga^2}{(y\cos V-\ga\sin V)^2+x^2\ga^2\cos^2 V}\,(dU^2+dV^2),\eeq
and it will be flat iff the denominator in the pre-factor is some constant. 
An easy algebraic discussion leads to the conclusion that the 
non-abelian model is flat iff
\beq\label{ff3}
\,x=\pm 1\,\quad \mbox{and}\quad y=0\quad\Leftrightarrow\quad 
R=\pm\,{\mb I}.\eeq 

\subsection{Frame geometry}
The spin connection, defined by 
\[\,de_a+\om_{ab}\wedge e_b=0,\]
is given by
\beq\label{fg3}
\om_{12}=\frac 1{2\ga}\,\frac{{\cal N}\,d\chi+xN\,d\tht}{x\De},\eeq
with
\beq\label{fg4}
{\cal N}=(a^2+\ga^2)'\,\De-(a^2+\ga^2)\,\De',\qq\quad N=a\,\De'-2a'\,\De,\eeq
where a prime indicates a derivative with respect to $\tht.$ One obtains
\beq\label{N}
N=2y\tht-2x\ga^2,\qq\qq \ga^2=x^2-y^2,\eeq
and
\beq\label{calN}
\frac{\cal N}{2x}=-y(\tht^2-\ga^2)+x(\ga^2-1)\,\tht.\eeq

The curvature (in the vielbein basis) has the single component
\beq\label{fg5}
R_{12,12}=\frac 1{\ga^2\De}\,\frac{({\cal N}\,\De'-\De\,{\cal N}')}{2x},\eeq
and we have for the Ricci
\beq\label{ricci}
{\rm Ric}\,_{11}={\rm Ric}\,_{22}=R_{12,12},\qq\qq {\rm Ric}\,_{12}=0.\eeq
The explicit form of the curvature component is 
\beq\label{fg7} 
\frac{{\cal N}\,\De'-\De\,{\cal N}'}{2x}=
x(\ga^2-1)(\tht^2-\ga^2)+4y\ga^2\tht.\eeq

\section{One-loop renormalizability}
The one-loop renormalizability is ensured if the Ricci tensor can be written
\beq\label{r1}
{\rm Ric}\,_{ij}=(\chi\cdot\pt)\,G_{ij}+\nabla_{(i}w_{j)},\qq 
\chi\cdot\pt=\sum\chi_k\,\frac{\pt}{\pt x_k},\eeq
where the $\{x_k\}$ are parameters of the lagrangian to be renormalized and $w$ is 
some (non-Killing) vector. In the vielbein basis this becomes
\beq\label{r2}\left\{\barr{l}
{\rm Ric}\,_{ab}= (e^{-1})_b^i\,(\chi\cdot\pt)e_{ai}+
(e^{-1})_a^i\,(\chi\cdot\pt)e_{bi}+{\cal D}_{(a}w_{b)},\\[4mm]
{\cal D}_a\,w_b=\wh{\pt}_a\,w_b+\om_{bs,a}\,w_s,\qq 
\wh{\pt}_a=(e^{-1})^i_a\,\pt_i.\earr\right.
\eeq
Out of these three relations, two involve only the unknown 
vector $w$:
\[-\frac 12\left({\cal D}_1\,w_2+{\cal D}_2\,w_1\right)=(e^{-1})_{a=1}^i\,
(\chi\cdot\pt)e_{b=2,i}+(e^{-1})_{a=2}^i\,(\chi\cdot\pt)e_{b=1,i},\]
\[-\frac 12\left({\cal D}_1\,w_1-{\cal D}_2\,w_2\right)=(e^{-1})_{a=1}^i\,
(\chi\cdot\pt)e_{b=1,i}-(e^{-1})_{a=2}^i\,(\chi\cdot\pt)e_{b=2,i}.\]
Defining
\beq\label{r3}
w_1= \sqrt{\frac x{\De}}\,\wh{w}_1,\qq\qq w_2=\sqrt{\frac x{\De}}\,\wh{w}_2,\eeq
we get
\beq\label{r4}
\wh{w}_1=\frac{\mu}{\ga}\,\tht,\qq\wh{w}_2=\frac{\mu}{\ga^2}(v+y\tht)+
(\tht+y/x)\,C_2-C_1,\eeq
where $\mu$ is a free parameter and 
\beq\label{r5}
C_1=2(\chi\cdot\pt)\left(\frac yx\right),\qq\qq 
C_2=\frac{x^2}{\ga^2}\,(\chi\cdot\pt)\,\left(\frac{\ga^2}{x^2}\right)\eeq
are constants.

Then we are left with the single relation
\beq\label{r6}
{\rm Ric}\,_{11}={\cal D}_1\,w_1+2\,(e^{-1})^i_{a=1}(\chi\cdot\pt)\,e_{a=1,i}.\eeq
The parameters to be renormalized are $x$ and $y$, so we define
\beq\label{r7}
\chi\cdot\pt=\chi_x\,\pt_x+\chi_y\,\pt_y.\eeq
The relation (\ref{r6}) gives only two relations (instead of three expected):
\beq\label{r8}\left\{\barr{l}
\ga^2\,\chi_x=xy\,\mu+x^2(\ga^2-1),\\[4mm]
-2y\,\chi_x+2x\,\chi_y=-x(\ga^2-1)\,\mu+4y\ga^2.\earr\right.\eeq
Since we are left with some freedom, let us see what comes out if we impose as 
a further condition the one-loop renormalizability of the T-dualized model. This 
is most conveniently done by switching to the parameters $\,u=x+y\,$ and $\,v=x-y.$ 
Then the transformation (\ref{nmet3}) becomes
\beq\label{r9}
u\ \ \to\ \ \wti{u}=\frac 1u,\qq\quad v\ \ \to\ \ \wti{v}=\frac 1v.\eeq
We will define for new renormalization constants
\beq\label{rr1}
\chi\cdot\pt=\wh{\chi}_u\cdot u\,\pt_u+\wh{\chi}_v\cdot v\,\pt_v,\qq\mbox{with}\qq 
\wh{\chi}_u=\frac{\chi_x+\chi_y}{x+y},\qq\wh{\chi}_v=\frac{\chi_x-\chi_y}{x-y}.
\eeq
Relation (\ref{r8}) implies
\beq\label{rr2}\left\{\barr{l}\dst 
\wh{\chi}_u(u,v)=+\frac{(1-v^2)}{2v}\,\mu(u,v)+2v\,\frac{u-v}{u+v}
+\frac{(u+v)(uv-1)}{2uv},\\[4mm]\dst 
\wh{\chi}_v(u,v)=-\frac{(1-u^2)}{2u}\,\mu(u,v)-2u\,\frac{u-v}{u+v}
+\frac{(u+v)(uv-1)}{2uv}.\earr\right.\eeq
Now $u$ (similarly for $v$) is renormalized by $\,u\,\wh{\chi}_u(u,v)\,$ while 
$\wti{u}$ is renormalized by $\,\wti{u}\wh{\chi}_u(\wti{u},\wti{v})\,$ so, 
the T-dual pair will be renormalizable provided that 
\beq\label{rr3}
\wh{\chi}_u( \wti{u},\wti{v})=-\wh{\chi}_u(u,v),\qq\quad
\wh{\chi}_v(\wti{u},\wti{v})=-\wh{\chi}_v(u,v).\eeq
So, if we define
\[m(u,v)=\frac{\mu(u,v)}{2}-\frac{u-v}{u+v},\]
the joint renormalizability constraints (\ref{rr3}) are equivalent to 
the single constraint
\beq\label{rr4}
m(\wti{u},\wti{v})=m(u,v).\eeq
The function $m$ describes the arbitrariness left over even after imposing the 
stability of renormalizability through T-dualization. 

We have therefore shown that the T-dual pair of Poisson-Lie $\si$-models is indeed 
renormalizable, in the strict field-theoretic sense, 
with the most general structure of the renormalization constants
\beq\label{rr5}\left\{\barr{l}\dst 
\frac{\mu}{2}=m+\frac{u-v}{u+v},\\[4mm]\dst 
\wh{\chi}_u=+m\,\frac{(1-v^2)}{v}+\frac{(1+v^2)(u-v)}{v(u+v)}
+\frac{(u+v)(uv-1)}{2uv},\\[4mm]
\dst \wh{\chi}_v=-m\,\frac{(1-u^2)}{u}-\frac{(1+u^2)(u-v)}{u(u+v)}
+\frac{(u+v)(uv-1)}{2uv},\earr\right.\eeq
where $\,m(u,v)\,$ is constrained by (\ref{rr4}). These renormalization constants 
are well defined since, as already stated in (\ref{riem}), the parameters 
$\,u+v=x\,$ and $\,uv=\ga^2\,$ never vanish.

\section{Conclusion}
Let us conclude with some remarks:
\brm
\item Absence of torsion and the dimension 2 for the target space is too poor to 
produce conformal geometries of interest for string theory. The case of 3d 
dimensional targets seem to be more promising from this point of view (cf. \cite{U}).
\item Remarkably, the renormalizability works despite the traceful structure 
constants of the Drinfeld double. However, it is important to note in this context 
that we work with strictly field theoretic renormalization where the Weyl mode 
is frozen to a constant value (the same is true for the  semi-abelian models). In the 
stringy framework, where the Weyl mode is coupled, the one-loop quantum equivalence 
requires that both sub-algebras ${\cal G}$ and $\wti{\cal G}$ in the Drinfeld double 
must have traceless structure constants \cite{bm}.
\item The problem of higher loop corrections to the Poisson-Lie T-duality appears to 
be more tricky than in the case of the Abelian or traditional non-Abelian T-duality. 
Of course, one problem to cope with is the fact that a finite one-loop renormalization 
can change the two-loops divergences \cite{bc}. But there is also a structural 
aspect of the thing: we expect that the two-loops effective action of the model 
{\em should not} probably have the same structure as the classical action (\ref{pl1}).
 Why? Because the Poisson-Lie symmetry should be itself only a semiclassical 
approximation of a quantum group symmetry and the full-fledged effective action 
should reflect the latter quantum symmetry rather than the former semiclassical one. 
This means that, starting from the two-loop level, we do not expect that the 
criterion of the strict field theoretical renormalizability should be respected 
but we rather believe that it should be replaced by a sort of quantum group Ward 
identities to be fulfilled by the effective action. It is only in the semiclassical 
(or one-loop) limit that these Ward identities would reduce to the strict requirement 
of the strict renormalizability. Unfortunately, we do not have any hint yet how 
to write down and test the hypothetical quantum group Ward identities. For the 
moment we just conclude that our one-loop analysis of all two-dimensional PLT targets 
indicates a good quantum health of the Poisson-Lie T-duality.
\erm


\begin{thebibliography}{333}
\bibitem{ks1} C. Klim\v c\'{\i}k and P. \v Severa, {\sl Phys. Lett.}  
{\bf B 351} (1995) 455;
\bibitem{K} C. Klim\v c\'{\i}k, {\sl Nucl. Phys.  Proc. Suppl.}, {\bf 46} (1996) 116;
\bibitem{ky} K. Kikkawa and M. Yamasaki, {\sl Phys. Lett.} {\bf B 149} (1984) 357;
\bibitem{ss} N. Sakai and I. Senda, {\sl Prog. Theor. Phys}, {\bf 75} (1986) 692;
\bibitem{ft} E. S. Fradkin and A. A. Tseytlin, {\sl Ann. Phys.}, {\bf 162} (1984) 31;
\bibitem{fj} B. E. Fridling and A. Jevicki, {\sl Phys. Lett.} {\bf B 134} (1984) 70;
\bibitem{oq} X. de la Ossa and F. Quevedo, {\sl Nucl. Phys.} {\bf B 403} (1993) 377;
\bibitem{ks2} C. Klim\v c\'{\i}k and P. \v Severa, {\sl Phys. Lett.}  
{\bf B 372} (1996) 65;
\bibitem{S1} K. Sfetsos, {\sl Nucl. Phys.} {\bf B 517} (1998) 549;
\bibitem{A} O. Alvarez, {Nucl. Phys. } {\bf B 584} (2000) 659;
\bibitem{cv} P. Y. Casteill and G. Valent, {\sl Nucl. Phys.} {\bf B 591} (2000) 491;
\bibitem{S2} K. Sfetsos, {\sl Phys. Lett.} {\bf B 432} (1998) 365;
\bibitem{bf} J. Balog, P. Forg\' acs, N. Mohammedi, L. Palla and J. Schnittger, 
{\sl Nucl. Phys.} {\bf B 535} (1998) 461.
\bibitem{hs1} L. Hlavat\'y and L \v Snobl, {\sl Mod. Phys. Lett.}  
{\bf A 17} (2002) 429;
\bibitem{jra} M. A. Jafarizadeh and A. Rezaei-Aghdam, {\sl Phys. Lett.} 
{\bf B 458} (1999) 477;
\bibitem{hs2} L. Hlavat\'y and L \v Snobl, {\sl Int. J. Mod. Phys.} {\bf A 17} 
(2002) 4043;
\bibitem{bc} G. Bonneau and P. Y. Casteill, {\sl Nucl. Phys.}  
{\bf B 607} (2001) 293;
\bibitem{bm} A. Bossard and N. Mohammedi, {\sl Nucl. Phys.} {\bf B 619} (2001) 128;
\bibitem{U} R. von Unge, JHEP {\bf 0207} (2002) 014.

\end{thebibliography}
\end{document}